\documentstyle [12pt] {article}

\newcommand{\ba} {{\bf a}}

\newcommand{\bj} {{\bf j}}
\newcommand{\bk} {{\bf k}}

\newcommand{\bu} {{\bf u}}
\newcommand{\bv} {{\bf v}}

\newcommand{\bx} {{\bf x}}

\newcommand{\bz} {{\bf z}}

\newcommand{\bB} {{\bf B}}

\newcommand{\bE} {{\bf E}}
\newcommand{\bF} {{\bf F}}

\newcommand{\bI} {{\bf I}}
\newcommand{\bJ} {{\bf J}}

\newcommand{\bP} {{\bf P}}
\newcommand{\bQ} {{\bf Q}}
\newcommand{\bR} {{\bf R}}

\newcommand{\Grad} {{\bf \nabla}}
\newcommand{\Div}  {{\bf \nabla} {\bf \cdot}}

\newcommand{\Curl} {{\bf \nabla} \times}

\newcommand{\pab}           {p_{ab}}
\newcommand{\qab}           {q_{ab}}
\newcommand{\rab}           {r_{ab}}
\newcommand{\tab}           {(1 - |p_{ab}|)}
\newcommand{\ea}            {{\bf e}_a}
\newcommand{\eb}            {{\bf e}_b}
\newcommand{\es}            {{\bf e}_\sigma}

\newcommand{\GOas}          {\Omega_a^\sigma}
\newcommand{\fas}           {f_a^{\sigma}}

\newcommand{\fab}           {f_{ab}}
\newcommand{\vab}           {\bu_{ab}}
\newcommand{\Bab}           {\bB_{ab}}
\newcommand{\vas}           {\bv_a^\sigma}

\newcommand{\Bas}           {\bB_a^\sigma}

\newcommand{\lab}[1]       {\label{#1}}

\newcommand{\la}           {\langle}
\newcommand{\ra}           {\rangle}

\newcommand{\r}            {\right}
\newcommand{\lf}            {\left}

\newcommand{\f} [2]	   {\frac{#1}{#2}}
\newcommand{\p} 	   {\partial}
\newcommand{\pf} [1]	   {\frac{\partial}{\partial {#1}}}
\newcommand{\pff} [1]	   {\frac{\partial^2}{\partial {#1}^2}}
\newcommand{\PD} [2] 	   {\frac {\partial {#1}} {\partial {#2}} }
\newcommand{\PDD} [2] 	   {\frac {\partial^2 {#1}} {\partial {#2}^2} }

\newcommand{\beq} {\begin{equation}}
\newcommand{\eeq} {\end{equation}}
\newcommand{\beqy} {\begin{eqnarray}}
\newcommand{\eeqy} {\end{eqnarray}}
\newcommand{\nn} {\nonumber}

\newcommand{\etal} {{\it et al. }}

\newcommand{\Ga} {{\alpha}}

\newcommand{\Gd} {{\delta}}
\newcommand{\GD} {{\Delta}}
\newcommand{\Ge} {{\epsilon}}

\newcommand{\Go} {{\omega}}
\newcommand{\GO} {{\Omega}}
\newcommand{\Gs} {{\sigma}}

\newcommand{\Gt} {{\tau}}

\begin{document}
\baselineskip=24pt
\bibliographystyle{unsrt}

\begin{center}
{\Large
\bf Lattice Boltzmann Magnetohydrodynamics}
\end{center}

\vspace{0.3in}

\begin{center}
Daniel O. Mart\'{\i}nez$^{1,2}$, Shiyi Chen$^{1,2}$
and William H. Matthaeus$^{2}$
\end{center}

\vspace{0.3in}

${}^{1}${\footnotesize Theoretical Division and Center for Nonlinear
Studies,
Los Alamos National Laboratory,} \\
\hspace*{0.3in}{\footnotesize Los Alamos, NM 87545}

${}^{2}${\footnotesize Bartol Research Institute, University of
Delaware, Newark, DE 19716}

\vspace{0.3in}

\date{  }
\begin{center}
{\bf Abstract}
\end{center}

Lattice gas and lattice Boltzmann methods are recently developed  numerical
 schemes for simulating a variety of physical systems.
In this paper a new lattice Boltzmann model for modeling
 two-dimensional incompressible magnetohydrodynamics (MHD) is presented. The
current model fully utilizes the flexibility of the lattice Boltzmann method
in comparison with previous lattice gas and lattice Boltzmann
 MHD models, reducing the number of moving directions from $36$
in other models to $12$ only. To increase computational efficiency, a
simple single time relaxation rule is used for
collisions, which
 directly controls the transport coefficients.
 The bi-directional streaming process of the particle distribution function
 in this paper is similar to the original model
[ H. Chen and  W. H. Matthaeus,
Phys. Rev. Lett., {\bf 58}, 1845(1987),
S.Chen, H.Chen, D.Mart\'{\i}nez and W.H.Matthaeus,
Phys. Rev. Lett. {\bf 67},3776 (1991)], but has been
greatly simplified, affording simpler implementation of boundary
conditions and
increasing the feasibility of
extension into a workable three-dimensional model.
Analytical expressions for the transport coefficients are presented.
Also, as example cases, numerical calculation for the Hartmann flow
is performed, showing a good agreement between the theoretical
prediction and numerical simulation, and a sheet-pinch simulation
is performed and compared with the results obtained with a spectral method.

\vspace{0.3in}


\vspace{0.3in}

\pagebreak
\section{Introduction}
Lattice gas automata (LGA) methods [1-5],
based upon dynamics of cellular automata (CA)
have attracted considerable
 attention during the last several years
for both modeling physical phenomena and simulating linear and
nonlinear partial differential equations.
The lattice gas method is
similar to traditional molecular dynamics in that
a particle representation is employed for microscopic processes
such as particle
collision and streaming, but the dynamics is much simpler.
The fundamental idea underlying the
lattice gas approach is
that simple microscopic dynamics may lead to macroscopic complexity.

The first lattice gas automata model was introduced by Frisch,
Hasslacher and Pomeau (FHP) \cite{fhp1} in a hexagonal lattice for
simulating two-dimensional (2D)
 hydrodynamics. The basic dynamical model
comprises particles
scattering and moving in discretized space and time.
The approach of this system is inspired by classical
statistical mechanics treatments of systems
such as the Ising model and simple cellular automata models.
Although intuitively appealing, the lattice automata method
for fluids requires an averaging
process in order to obtain the macroscopic
fluid variables and their dynamics,
due to the high levels of noise naturally present in the
discretized particle representation.
More recently there has been a trend towards
using the lattice Boltzmann (LB) scheme instead of the
lattice gas automata method.
Unlike the lattice gas method in which one keeps track of each individual
particle, in the lattice Boltzmann approach we are interested only in
the one-point distribution function.
 While retaining the advantages associated
 with parallel implementation of lattice automata,
the LB method is more efficient and accurate
computationally, and essentially noise-free.

The history of lattice gas and
lattice Boltzmann models for magnetohydrodynamics (MHD) can be very briefly
summarized as follows. The first attempt to model 2D MHD with a lattice
gas automata scheme was
 carried out by Montgomery and Doolen \cite{mondool1,mondool2}.
In their model the basic FHP
model is extended to include
additional degrees of freedom to account for the vector potential.
 To update the dynamics,
 some space average quantities need to be evaluated.
By doing so, the essential feature of locality, that characterizes lattice
gas systems, is lost. In addition, because of  the vector potential
representation
of magnetic field the model is intrinsically two dimensional. It is noted that
the recent lattice Boltzmann MHD model by Succi \etal
 \cite{succi} has
 a similar limitation.
 Another MHD lattice gas automata model with pure local operations
 was proposed by H. Chen \etal \cite{hudong-bill,hudong-bill2}.
To account for the Lorentz force,
the model introduced a tensor (i.e., two-indexed) particle
representation, and
a random walk (or, ``bi-directional streaming'') mechanism.
For each one of these particles, there are
two vectors attached,
representing the momentum and magnetic field vectors.
During the streaming procedure, particles move along one of the two
possible vector
directions with a probability deduced by requiring
that MHD behavior is obtained macroscopically. Later, S. Chen \etal \cite{ccmm}
extended the lattice gas automata
model into a lattice Boltzmann model. The simulation of the
two dimensional LGA
and LB models for problems with  free boundary and simple wall boundaries,
including the two dimensional Hartmann flow and two dimensional magnetic
reconnection, has achieved reasonable success \cite{ccmm,mhdcasyc} in
test problems.

The random walk MHD LGA model
\cite{hudong-bill,hudong-bill2}, and its extension to the
LB scheme \cite{ccmm}, however, have two major problems.
First, because of the random streaming,
implementation of wall boundary conditions becomes complicated,
requiring increased
 computational memory and computational work \cite{ccmm}.
Second and most important, although both the lattice gas automata model
\cite{hudong-bill,hudong-bill2} and the lattice
Boltzmann model \cite{ccmm} can be formally
extended
into three dimensional (3D) space,
real 3D implementation is impractical due
to
memory requirements. In order to
include a correct Lorentz force, for a lattice with $N$ moving directions,
$N\times N$ particle states are needed.
In 3D LGA and LB models,
a face-centered-hyper-cubic
(FCHC) lattice of $N = 24$ is usually employed.
For this case, the random walk model needs at least $576$ states,
 requiring about 1.2 gigabytes for a system of $64^3$. Thus,
to some degree, its actual value as a computational tool is diminished
because of the requirement of vast amounts of memory.

In the present paper we introduce a new lattice Boltzmann model
for MHD that requires considerably less memory than previous models,
while continuing to offer the computational efficiency of the LB
approach. The new model is, in essence, a reduction of the previous
 MHD LBE model \cite{ccmm}, including a smaller number of allowed
states, while maintaining the symmetries and most of the
desirable analytical and computational properties of the earlier
method. The new model utilizes a ``13 bit'' representation on a 2D
hexagonal lattice, in contrast to the earlier requirement of a
37-bit 2D model.

The paper is organized as follows:
In Section II we will describe the model and show
how ideal MHD
is obtained in the fluid limit.
In Section III
the next order terms in the Chapman-Enskog expansion are included,
adding dissipative effects (viscosity and resistivity) to the model.
Following this, two sections are devoted to numerical tests.
Section IV discusses the linear Hartmann flow problem as a first
test of the model, for several Hartmann numbers $H$,
which parameterizes the solutions.
Section V describes use of the model
to solve numerically for the evolution of the
MHD sheet-pinch configuration, i.e. the dynamics
of a highly sheared planar
magnetic field.
The results obtained with the LB run are compared with those obtained
with a spectral method run, using the same initial conditions.
Discussions of the results and of the model are presented in Section
VI.
Finally,
in Appendix A some useful tensorial relations for the derivation of
the model are presented for completeness;
and in Appendix B we show how the same principle that is applied
to obtain MHD behavior on the hexagonal lattice, can be extended to
the square lattice.

\section{Description of the Model}
The model described here is
inspired by the previous random walk MHD LGA (or, CA) model
\cite{hudong-bill,hudong-bill2}
and is motivated by the
need for an MHD LB scheme that
is computationally feasible. This requires
overcoming the two problems mentioned above.
The model in this
section, for simplicity, concerns two dimensional problems, but its
principle also applies to three-dimensional models.

For our two-dimensional system, we use the standard hexagonal grid \cite{fhp1}.
In the vicinity of given lattice point at $\bx$,
six nearest neighbors are located at positions
$\bx + \ea$, with $\ea=(cos(2 \pi(a-1)/6),sin(2 \pi(a-1)/6))$,
$a = 1, \dots,6$. Instead of using $6\times 6$ particle states as in
previous models, we only consider a subset of them. Each state
is labeled by the pair of indices $(a,\Gs)$. The positive particle
distribution function is represented by $f_a^\Gs$ with
$a=1,\dots,6$ and $\Gs=1,2$, where $\Gs$ is defined relative to $\ea$
in the following manner; $\Gs=1$ corresponds to the direction
$a+1$ (mod 6), and $\Gs=2$ to $a-1$ (mod 6).

The evolution of the system consists of a sequence
of a streaming stage in which the distribution $\fas$ is propagated
from each cell to its neighbour cells, followed by a collision
stage in which the distribution at each cell is
redistributed according to some conservation laws, as we will see below.
The propagation part
of the evolution for our particular model
consists of partitioning the particle distribution into the two
directions associated with the state ($a,\Gs$),
\beq
f_a^\Gs({\bf x},T) \rightarrow (1-p) f_a^\Gs({\bf x} +{\bf e}_a ,T+1)
+ p f_a^\Gs({\bf x} +\es ,T+1),  \lab{3.54}
\eeq
where $T$ corresponds to the discrete microscopic time and $p$ is a given
parameter which represents the fraction of the distribution
function $\fas$ that propagates along the $\Gs$
direction. Notice that this streaming procedure improves in two ways
the random walk used in \cite{hudong-bill,hudong-bill2,ccmm}. First,
the motion of the ``magnetic'' portion of the distribution function
$f_a^\Gs$ is always ``forward'' because the streaming parameter $p$ is
greater than zero. (For the $36$-bit model \cite{ccmm}, the
distribution function is represented by $f^b_a$, ($a=1,\dots 6$,
$b=1,\dots 6$) and a fragment
of $f_a^b$ moves in the direction $sign(\pab) \eb$ while
the remainder of the distribution moves in the direction $\ea$. Therefore,
there are states $(a,b)$ for which the distribution streams in the
direction $-\eb$).
In addition, in the present model
the angle between the two directions $\ea$ and
$\es$ is $\pi/3$ for all states $(a,\Gs)$. These properties are important
for imposing boundary conditions.
In addition to these twelve states,
the model includes a $13^{th}$ state denoted by
$f_0$ that represents the fraction of the distribution function at a cell that
does not advect at all. This ``stopped'' distribution introduces additional
freedom in the model that allows to get rid of undesirable dependence
of the pressure on the velocity that had plagued earlier LGA and LB models
\cite{chen-chen-bill,qian}.

Associated with each state $(a,\Gs)$ we define the local
microscopic velocity $\bv_a^\Gs$, which is equal to the mean
velocity at each cell, and the microscopic magnetic field $\bB_a^\Gs$
\beqy
\vas  &\equiv& (1-p) \ea + p \es, \\
\bB_a^\Gs &\equiv&     r \ea + q \es.
\eeqy
Although in principle the
parameters $r$, $q$ and $p$ are unrelated, we will see later that a
connection between $r$ and $q$ is set by the dynamical requirements.
Notice that unlike the velocity, the microscopic magnetic field does
not, on the surface, appear to play an active role in the evolution of
the system. (However, later we will see that this is not the case;
see the discussion after Eq. (\ref{induction}).)

The density $\rho$ is defined as the summation of all particle
distribution functions,
\beq
\rho(\bx,T) \equiv f_0(\bx,T) + \sum_{a=1}^6 \sum_{\Gs=1}^2
f_a^\Gs (\bx,T).        \label{density}
\eeq
In (\ref{density}) the limits of summation on $a$ and $\sigma$ are given
explicitly; subsequently we will
sometimes suppress them for convenience when no ambiguity is
introduced.

The macroscopic velocity and magnetic field are defined as averages
of the microscopic fields $\bv_a^\Gs$ and $\bB_a^\Gs$,
\beqy
\rho \bv \equiv \sum_{a,\sigma} \bv_a^\Gs f_a^\Gs \label{velocity},\\
\rho \bB \equiv \sum_{a,\sigma} \bB_a^\Gs f_a^\Gs \label{magnetic}.
\eeqy

The
kinetic equation obeyed by $f_a^\Gs$ can be written
by combining the effect of streaming, represented
by (1) with the collisional effects, denoted by
the symbol $\GO_a^\Gs$, arriving at,
\begin{eqnarray}
f_a^\Gs (\bx,T) = (1-p) [f_a^\Gs (\bx -\ea,T-1) +
\Omega_a^\Gs(\bx - \ea,T-1) ] + \nn \\
p [f_a^\Gs (\bx - \es,T-1) + \Omega_a^\Gs(\bx -
\es,T-1)].  \label{kineticdiscrete}
\end{eqnarray}

\noindent
In the present model, for simplicity,
we assume that the collision operator $\Omega_a^\Gs$
has a single time relaxation form \cite{ccmm},
\[  \Omega_a^\Gs = -(f_a^\Gs - f_a^{\Gs(eq)})/ \tau,  \]
where $\tau$ is the relaxation time and $ f_a^{\Gs(eq)}$ is the local
equilibrium distribution function depending on the local
particle density, velocity
and magnetic field.

A crucial step in the development of a LB method is the
selection of an appropriate single particle
equilibrium distribution function, associated with vanishing of the
collision operator. This equilibrium distribution function
has to be consistent with definitions (\ref{density}),
(\ref{velocity}) and (\ref{magnetic}), and in addition has to give
rise to the MHD equations.
A suitable equilibrium distribution function fulfilling all these
conditions
\protect\label{fas} is given by
\beqy
f_a^{\Gs(eq)} &=&
 (\f{\rho}{12}) \lf\{ \f{12}{\alpha +12} + \r.
\f{4}{C} \lf[ \f{}{}\vas \cdot \bv + \r.
\f{(2-p)^2}{3 q^2} \Bas \cdot \bB  \nn\\
&+& \f{4(2p-1)}{C} [(\es \cdot \bv)^2 -
 (\es \cdot \bB)^2 ]   \nn \\
&+& \f{4(1-p^2)}{C}[ (\ea \cdot \bv)
        (\es \cdot \bv)
  -   (\ea \cdot \bB)(\es \cdot \bB)]         \nn \\
&+& \f{2(2-p)}{3 q} [(\ea \cdot \bv)(\es \cdot \bB)
-  (\ea \cdot \bB)(\es \cdot \bv)]  \nn\\
&-& \f{(2p -1)(2 - p)}{C} v^2 - \f{p^2 - 4p + 1}{C}B^2
\lf.\lf.\f{}{}\r] \r\},  \lab{faseq}
\eeqy
and
\beq
f_0^{(eq)} = \rho \lf[ \f{\Ga}{12 + \Ga} - \f{2}{C} v^2 \r], \lab{f0seq}
 \eeq
where $C= 2 (p^2 - p +1)$.
Positivity of the distribution function is guaranteed if $v$ and $B$ are
sufficiently small.

To derive the MHD fluid model, we next
form the continuum kinetic equation by Taylor
expanding (\ref{kineticdiscrete}) in the limit of low
frequencies and
long wavelengths \cite{fhp1}. The result to lowest order is
\begin{equation}
\f{\p f_a^\Gs}{\p t} + \vas \cdot \Grad f_a^\Gs = \Omega_a^\Gs.
\label{kineticcont}
\end{equation}
Equations for the density, momentum transport and magnetic momentum
transport can now be found by taking moments of (\ref{kineticcont}):
\beqy
&&\f{\p \rho}{\p t} + \Grad \cdot (\rho \bv) = 0, \nn \\
&&\f{\p}{\p t} (\rho \bv) + \Div {\bf \Pi}^{(0)} = 0, \label{transport} \\
&&\f{\p}{\p t} (\rho \bB)+ \Div {\bf \Lambda}^{(0)} = 0,  \nn
\eeqy
where the momentum flux tensor ${\bf \Pi}^{(0)}= \sum_{a,\Gs} \vas
\vas f_a^{\Gs(eq)}$, and the magnetic momentum flux tensor
${\bf \Lambda}^{(0)}= \sum_{a,\Gs} \bB_a^\Gs \vas f_a^{\Gs(eq)}$
are defined in a similar fashion as in the $36$-bit model
\cite{hudong-bill2,ccmm}.
Plugging the expression for $f_a^{\Gs(eq)}$, Eqs. (\ref{faseq}) and
(\ref{f0seq}),
in the above definitions,
we may compute the fluxes of density, momentum and magnetic field.

In computing the flux tensors, we need to make use of some of the
freedom contained in our definition of the equilibrium distribution
(\ref{faseq}) and (\ref{f0seq}), which depends on parameters $p$, $q$, $r$, and
$\alpha$. This flexibility
in selecting the parameters
will be used to eliminate certain unphysical terms that would prevent
the appearance of MHD equations at leading order, and also to
obtain other desirable properties.

First, we note that the presence of
terms that mix the directions ${\bf e}_a$ and $\es$ appears to be
necessary to obtain a correctly structured induction equation.
Next, the structure of the equilibrium permits
appearance of an unphysical pressure-like term in the
induction equation. This can be eliminated
by choice of a relationship between $r$ and $q$, namely,
\beq
  r = -q \f{1+p}{2-p}.
\label{rqcond}
\eeq

Now we turn to some considerations with regard to selection of the
value of the parameter $p$.
Two interesting properties of this MHD system should be mentioned. First,
in the limit case of $p=0$, for which the streaming is only along the
$a$ direction in an $(a,\Gs)$ state, incompressible hydrodynamics is
recovered, as expected.
Second, if the streaming parameter $p$
is changed to $1-p$, the roles played by $a$ and $\Gs$ are interchanged.
To clarify this point, let us insert the relation between $r$ and $q$
into $\Bas$ to obtain
\beq
\bB_a^\Gs = \f{q}{2-p} [-(1+p) \ea + (2-p) \es ],  \nn
\eeq
and recall that $\vas = (1-p)\, \ea + p\, \es$.
If $p$ is replaced by $1-p$, then $2-p$ and $1+p$ (and thus $q$ and
$r$) also interchange their values. This property holds
for the distribution function $f_a^\Gs$ as well; in
other words, the same macroscopic MHD properties are obtained under
this exchange, including sound speed, and
transport coefficients.
The streaming parameter $p$ is constrained to be between
zero and one
for mass conservation. However, we will see
in the next section that choosing $p=1/2$
eliminates spurious terms that would otherwise
appear in the momentum and
induction equations at second order in the spatial expansion
of the kinetic equation.

Turning to the parameter $q$, we note that,
unlike $p$, $q$ can in principle take
any desired real value with the exception of zero. For
simplicity, we chose $q = (2-p)/\sqrt{3}$. For this particular
value of $q$,  $|{\bf v}_a^\Gs|=|{\bf B}_a^\Gs|$ and the microscopic
velocity and magnetic field have the same intensity.
Magnetohydrodynamics behavior is obtained for either
$q$ positive or negative. Notice that changing the
sign of $q$ will reverse the direction of $\bB$. Therefore,
this feature is associated to the fact that if the magnetic field
is reversed everywhere the fluid flow is unchanged.
This property was already present in the $36$-bit MHD CA
model \cite{hudong-bill2}) and in its LB version \cite{ccmm}.

The equilibrium distribution function for these values of $p$
and $q$ becomes,
\beqy
f_a^{\Gs(eq)} &=&
 (\f{\rho}{12}) \lf\{ \f{12}{\alpha +12} + \r.
\f{8}{3} \lf[ \f{}{}\vas \cdot \bv + \r. \Bas \cdot \bB  \nn\\
&+& 2[ (\ea \cdot \bv)(\es \cdot \bv)
  -   (\ea \cdot \bB)(\es \cdot \bB)]         \nn \\
&+& \f{2}{\sqrt{3}} [(\ea \cdot \bv)(\es \cdot \bB)
-(\ea \cdot \bB)(\es \cdot \bv)] + \f{B^2}{2}\lf.\lf.\f{}{}\r] \r\},
\eeqy
and
\beq
f_0^{(eq)} = \rho \lf[ \f{\Ga}{12 + \Ga} - \f{4}{3} v^2 \r].
\eeq
Using this form of the equilibrium, after some straightforward algebra
explicit
forms for the flux tensors are obtained as,
\beqy
& &\Pi_{ij}^{(0)} = \f{\rho}{2} \left\{\f{9}{12 + \Ga} \Gd_{ij} +
 u_k u_l [ \Delta_{ijkl} - \Gd_{ij} \Gd_{kl}] -
 B_k B_l [ \Delta_{ijkl} - 2 \Gd_{ij} \Gd_{kl}] \right\}, \\
& &\Lambda_{ij}^{(0)} = \rho\,(\Gd_{il} \Gd_{kj} - \Gd_{ik} \Gd_{jl}) u_k
B_l.
\label{fluxtensors}
\eeqy
The ideal MHD equations emerge from this procedure,
\beqy
&&\rho \PD{\bv}{t} + \rho (\bv \cdot \Grad) \bv  = -\Grad(P + B^2/2) +
(\bB \cdot \Grad) \bB \label{ns} \\
&&\PD{\bB}{t} + (\bv\cdot\Grad)\bB = (\bB\cdot\bf\Grad) \bv,
\label{induction}
\eeqy
where the pressure $P=\rho C_s^2$. $C_s$ is the sound speed of the
system and has a simple form related to $C$ and $\alpha$,
$C_s = \sqrt{3 C / (12 + \alpha)}$, thus being controllable
by the parameters $p$ and $\alpha$.

Now we make a small digression to clarify some points about
the microscopic properties of the $13$-bit model.
For the $36$-bit MHD model (LGA or LBE), we recall that the macroscopic
velocity and magnetic field are defined as follows
\cite{hudong-bill,hudong-bill2,ccmm},
\beqy
\rho \bv&=&\sum_{a,b}^6 \vab \fab=
\sum_{a,b}^6 \lf[ \tab \ea + \pab \eb \r] \fab \nn\\
\rho \bB&=&\sum_{a,b}^6 \Bab \fab=
\sum_{a,b}^6 \lf[ \rab \ea + \qab \eb \r] \fab.\nn
\eeqy
The parameter matrices $\bP$, $\bQ$, and $\bR$ (of elements
$\pab$, $\qab$, and $\rab$, respectively)
involve, in principle, $108$ independent scalars.
Arguing that some conditions must be placed
on these matrices to obtain the right MHD behavior, it was
possible
to reduce the number of independent scalars to only
six \cite{hudong-bill2}. Although we have not
explicitly imposed such conditions for the case of the $13$-bit model, we
want to show that these properties are already present in the
model as it is.

For the case of the $36$-bit model, because the microphysics should be
isotropic, we expect that rotating $\ea$ and $\eb$ together by a multiple
of $\pi/3$, $\vab$ and $\Bab$ should rotate by the
same amount. This
condition implies that the matrices should be circulant.
Recall that a tensor $\Xi$ is circulant if for $c=1,\cdots,6$,
$\Xi_{ab}=\Xi_{a+c,b+c;(mod\,6)}$, $a,b=1,\cdots,6$.
In addition, the microscopic physics should be mirror symmetric:
if $\ea$ and $\eb$ are interchanged, then the new values for $\vab$ and
$\Bab$, should be the mirror images with
respect to the line that bisects the
angle between $\ea$ and $\eb$. This implies
that the matrices are symmetric,
$\Xi_{ab}=\Xi_{ba}$.
These two conditions are trivially obeyed by our $13$-bit
model because the matrices  $\bP$, $\bQ$, and $\bR$ in the
$36$-bit model are the counterpart of the scalars $p$, $q$, and $r$.

There is another constraint to be enforced on the
coefficients that is
more subtle, and is associated with the vector nature of the
velocity, and pseudovector nature of the magnetic field.
There exist microscopic transformations that  reverse the direction
of one of the fields ($\bv$ or $\bB$) everywhere, while leaving
the other one unchanged. Imposing such a property guarantees,
for example that if $\bB\rightarrow-\bB$ the evolution of the
velocity field will remain unchanged, while $-\bB$ becomes the
solution for the induction equation.
This property was also included in the $36$-bit model,
by imposing some constraints on the parameter matrices,
namely, $\pab=-p_{ab+3}$,
$\qab=q_{ab+3}$, and $\rab=-r_{ab+3}$, where all the sums
are {\it modulo}$(6)$. It can be easily seen from the
definitions $\vab=\tab\ea + \pab\eb$ and $\Bab=\rab\ea+\qab\eb$ for the
$36$-bit model, that
by changing every $b$ by $b+3$ $\vab$ is reversed, whereas $\Bab$
is unchanged. The opposite is true if every  $a$ is replaced by $a+3$. In
summary,
\beqy
&&{\bf B}_{ab+3} = \Bab  \hspace{5.7em} {\bf u}_{ab+3} = -\vab \nn \\
&&{\bf B}_{a+3b} = -\Bab  \hspace{5em} {\bf u}_{a+3b} = \vab. \nn
\eeqy
This property should also be obeyed by the $13$-bit model we are
dealing with in
this section. There is no straightforward
``translation'' from the
$a+3$ or $b+3$ operation in the $36$-bit scheme to
our model here, because now for every
$a$ there are only two $b$'s: $a+1$ and $a-1$ $modulo(6)$.
Nevertheless, we find that the $13$-bit model obeys the following
relationships,
\beqy
&&{\bf B}_{a+4,2} = {\bf B}_{a1} \hspace{5.7em}
{\bf u}_{a+4,2} = - {\bf u}_{a1}\nn \\
&&{\bf B}_{a+1,2} = -{\bf B}_{a1}  \hspace{5em}
{\bf u}_{a+1,2} = {\bf u}_{a1}, \nn
\eeqy
and therefore, such transformations are already embedded in the model
as it is.

As a final comment, we note that this scheme would not exhibit
MHD behavior in the context of a CA-type lattice gas model.
If the idea of the
splitting of the distribution in two parts of the present
model is ``translated'' to the CA realm, we would, most likely,
end up with the $36$-bit lattice gas scheme of Refs.
\cite{hudong-bill,hudong-bill2} that
inspired our model, in the first place. Although this final
statement would be hard to rigorously prove,
we suspect that the present lattice Boltzmann scheme allowed us to get
rid of all the ``degeneracies'' (or most of them) present in the $36$
bit model.
\section{Transport Coefficients}
In this section we examine in detail the structure of the present
model from the perspective of
a Chapman-Enskog expansion procedure.
This renders the long wavelength, low frequency
behavior of the system, including corrections to the ideal equations
in the form of dissipative transport effects.
We start from the discrete kinetic equation
(\ref{kineticdiscrete})
\beqy
f_a^\Gs (\bx,T) = (1-p) [f_a^\Gs (\bx -\ea,T-1) +
\Omega_a^\Gs(\bx - \ea,T-1) ] + \nn \\
p [f_a^\Gs (\bx - \es,T-1) +
\Omega_a^\Gs(\bx - \es,T-1)], \nn
\eeqy
 and expand up to second order in time
and space variables, to obtain
\beqy
&&\PD{\fas}{t} + \vas \cdot \Grad(\fas + \GOas) -
\vas \cdot \Grad \PD{(\fas+\GOas)}{t} - \GOas
-\f{1}{2}\f{\p^2\GOas}{\p t^2}\nn \\
&&- \f{1}{2}[(1-p)\ea \ea + p \es \es]:\Grad \Grad (\fas + \GOas) +
\PD{\GOas}{t} -\f{1}{2} \PDD{\fas}{t} = 0 \label{kin-exp}.
\eeqy
Now we adopt the following multiple scale expansion \cite{fhp2,sycthermo}.
The time derivative is expanded as
\beq
\pf{t} = \Ge \pf{t_1} + \Ge^2 \pf{t_2} + \cdots,
\eeq
where $\Ge$ is the expansion parameter assumed small, implying
that $t_2$ is a slower time scale than $t_1$, and will be
associated with diffusion effects.
Likewise, the one-particle distribution function is expanded, assuming
small departures from equilibrium,
\beq
\fas = f_a^{\Gs(0)} + \Ge f_a^{\Gs(1)} + \Ge^2 f_a^{\Gs(2)},
\eeq
where $f_a^{\Gs(0)} = f_a^{\Gs(eq)}$.
Finally, for the collision operator we write
\beq
\GOas = -\f{1}{\tau} (\Ge f_a^{\Gs(1)} + \Ge^2 f_a^{\Gs(2)}).
\eeq
Replacing all these expansions into (\ref{kin-exp}), we find the following
relation to order $\Ge$,
\beq
\PD{f_a^{\Gs(0)}}{t_1} + \vas \cdot \Grad f_a^{\Gs(0)} =
-\f{f_a^{\Gs(1)}}{\Gt},
\label{ordere1}
\eeq
and
\beqy
&&\PD{f_a^{\Gs(1)}}{t_1} + \PD{f_a^{\Gs(0)}}{t_2} +
\vas \cdot \Grad f_a^{\Gs(1)}
- \vas \cdot \Grad \pf{t_1} f_a^{\Gs(0)} \nn\\
&&\hspace{3em}- \f{1}{2} [(1-p)\ea \ea + p \es \es]:\Grad \Grad f_a^{\Gs(0)}
\nn\\
&&\hspace{5em}-\f{1}{\Gt}\pf{t_1} f_a^{\Gs(1)}
-\f{1}{2} \pff{t_1} f_a^{\Gs(0)} =
-\f{1}{\Gt} f_a^{\Gs(2)} \label{ordere2}
\eeqy
to order $\Ge^2$. From (\ref{ordere1}) we can obtain the auxiliary
relationship
$$
\f{1}{2\Gt}(\pf{t_1} + \vas \cdot \Grad) f_a^{\Gs(1)} =
-\f{1}{2} \left[ \PDD{f_a^{\Gs(0)}}{t_1} +
2 \vas \cdot \Grad \PD{f_a^{\Gs(0)}}{t_1} +
\vas \vas : \Grad \Grad f_a^{\Gs(0)} \right]
$$
that can be combined with (\ref{ordere2}), and after some
algebraic manipulations the following equation is obtained:
\beqy
\PD{ f_a^{\Gs(0)} }{t_2} + (1 - \f{1}{2\Gt})
\lf[\PD{f_a^{\Gs(1)}}{t_1} +  (\vas \cdot \Grad) f_a^{\Gs(1)}\r] &=&
 \nn\\
\hspace{4em}
\f{p(1-p)}{2} (\ea - \es)(\ea - \es):\Grad \Grad f_a^{\Gs(0)} &=&
-\f{1}{\Gt} f_a^{\Gs(2)} \label{ordere2b}.
\eeqy
Summing equations (\ref{ordere1}) and (\ref{ordere2b})
over all velocities, and using that \\
\mbox{$\sum_{a,\Gs} f_a^{\Gs(1)}=0$},
and $\sum_{a,\Gs} f_a^{\Gs(2)}=0$, the following continuity equation up to
second order is obtained,
\beq
\PD{\rho}{t} + \Div(\rho \bv) =
\Ge^2 \f{p(1-p)}{2}
\sum_{a\Gs} (\ea - \es)(\ea - \es) :\Grad \Grad f_a^{\Gs(0)}
\label{continuity13}.
\eeq
The term on the right hand side can be calculated using the tensorial
relationships (\ref{tensors}) in Appendix A,
and the expression for $f_a^{\Gs(eq)}$,
\beqy
\PD{\rho}{t} + \Div(\rho \bv) =
\Ge^2 \, \f{p(1-p)}{2} \f{\p}{\p x_i} \left\{ \left(\f{6}{12+\Ga}\right)
\f{\p \rho}{\p x_i} +
\f{3}{C^2} \f{\p (\rho v^2)}{\p x_i}  \right. \nn \\
+ \f{2 p^2-2 p -1}{C^2} \left[
2 \f{\p (\rho u_i u_j)}{\p x_j} +
 \f{\p (\rho B^2)}{\p x_i} -
 \left. 2 \f{\p (\rho B_i B_j)}{\p x_j} \right] \right\}.
\eeqy
Therefore, there are second order corrections to the continuity equation.
The most important term is apparently the one associated
with the density diffusion, compared to the other terms that
are quadratic in the fields $u$ and $B$.
In the limit of hydrodynamics, i.e. $p=0$ these additional terms vanish.
No value of $0<p<1$ (necessary for mass conservation)
will make these spurious terms vanish. However, notice that
the r.h.s. vanishes when integrated over the whole domain,
and mass conservation is restored.

Similarly, adding moments of equations (\ref{ordere1}) and
(\ref{ordere2b}) with respect to $\vas$ and $\Bas$, the following
momentum equation and induction equation are obtained:
\beqy
\PD{(\rho \bv)}{t} + \Div {\bf \Pi} =
 \,\f{p(1-p)}{2}
\sum_{a\Gs} \vas (\ea - \es)(\ea - \es) :\Grad \Grad f_a^{\Gs(0)},
\lab{110}    \\
\PD{(\rho \bB)}{t} + \Div {\bf \Lambda} =
 \,\f{p(1-p)}{2}
\sum_{a\Gs} \Bas (\ea - \es)(\ea - \es) :\Grad \Grad f_a^{\Gs(0)},
\lab{111}
\eeqy
where
$$
{\bf \Pi} = \sum_{a\Gs} \vas \vas \left[ f_a^{\Gs(0)}
+ \Ge (1-\f{1}{2\Gt}) f_a^{\Gs(1)}\right],
$$
and
$$
{\bf \Lambda} =
\sum_{a\Gs} \Bas \vas \left[ f_a^{\Gs(0)}
+ \Ge (1-\f{1}{2\Gt}) f_a^{\Gs(1)}\right].
$$
We can see that there will be several contributions to the transport
coefficients coming from (\ref{110}) and (\ref{111}).
Let us first examine the contribution coming from the right hand side
of the equations. For both the momentum and the induction equations
only the terms linear in the fields in $f_a^{\Gs(0)}$ will be
different from zero, due to the microscopic symmetries already embedded
in the model at microscopic level. From the previous section
we can write
\beq
f_a^{\Gs(0)} \sim \f{\rho}{12 + \Ga} + \f{\rho}{3 C}
[\vas \cdot \bv + \Bas \cdot \bB ] + O(v^2,B^2).
\eeq
The following tensorial relationships are needed first,
\beqy
& &\sum_{a\Gs} (\vas)_i(\ea - \es)_j(\ea - \es)_k(\vas)_l =
\f{3}{4}(C -3) (\Gd_{ij} \Gd_{kl} + \Gd_{ik} \Gd_{jl}) +
\f{3}{4}(C + 3)\,\Gd_{il} \Gd_{jk} \nn \\
& &\sum_{a\Gs} (\vas)_i(\ea - \es)_j(\ea - \es)_k(\Bas)_l =
\f{3 \sqrt{3}}{4} (2p-1) (\Gd_{ij} \Gd_{kl} + \Gd_{ik} \Gd_{jl} -
\Gd_{il} \Gd_{jk} ) \\
& & \sum_{a\Gs} (\Bas)_i(\ea - \es)_j(\ea - \es)_k(\Bas)_l =
-\f{3}{4} (C -3) (\Gd_{ij} \Gd_{kl} + \nn \\
& & \hspace{2cm}\Gd_{ik} \Gd_{jl}) +
\f{9}{4} (C -1) \Gd_{il} \Gd_{jk}. \nn
\eeqy
Using the above relationships and after some algebra, we obtain
\beqy
&&\f{p(1-p)}{2}
\sum_{a\Gs} \vas (\ea - \es)(\ea - \es) :\Grad \Grad f_a^{\Gs(0)} =
\nn\\
&&\hspace{3em}
\f{p(1-p)}{8 C} \lf[ 2(C-3) \Grad(\Div(\rho\bv))
 + (C+3) \nabla^2 (\rho\bv) \right] \nn\\
&&\hspace{3em}
+\sqrt{3}(2p-1) \left[ 2 \Grad(\Div(\rho\bB)) -
\nabla^2(\rho\bB)\right],
\eeqy
and
\beqy
&&\f{p(1-p)}{2}
\sum_{a\Gs} \Bas (\ea - \es)(\ea - \es) :\Grad
\Grad f_a^{\Gs(0)}
\nn\\
&&=\hspace{2em}
\f{\sqrt{3}p(1-p)}{8 C} (2p-1) \left[ 2\Grad(\Div(\rho\bv)) -
 \nabla^2 (\rho\bv) \r] \nn\\
&&\hspace{3em}
+ \f{1}{\sqrt{3}} \lf[ -2(C-3) \Grad(\Div(\rho\bB)) +
3(C -1) \nabla^2(\rho\bB)\r].
\eeqy
The other contribution, that is controllable through $\Gt$, comes from
\beqy
\Pi_{ij}^{(1)} &=& (1-\f{1}{2\Gt})
\sum_{a\Gs} (\vas)_i (\vas)_j f_a^{\Gs(1)},\\
\Lambda_{ij}^{(1)} &=& (1-\f{1}{2\Gt})
\sum_{a\Gs} (\Bas)_i (\vas)_j f_a^{\Gs(1)},
\eeqy
with $f_a^{\Gs(1)}=-\Gt \left[ \p/\p t +
\vas \cdot \Grad \right]f_a^{\Gs(0)}$.
We obtain as a contribution to the momentum equation,
\beq
\Div {\bf \Pi}^{(1)} = (\Gt-\f{1}{2})
\left\{ (\f{C}{4} + C_s^2) \Grad[\Div(\rho\bv)] +
\f{C}{8} \Grad^2(\rho \bv) \right\},
\eeq
whereas the contribution to the induction equation will be
\beq
\Div {\bf \Lambda}^{(1)} = (\Gt-\f{1}{2}) \f{C}{4}
\left\{ -\Grad[\Div(\rho\bB)] + \f{3}{2} \Grad^2(\rho \bB) \right\}.
\eeq
We can write the macroscopic equations obtained, including all
the above contributions to the transport coefficients and in the limit
of low Mach number:
\beqy
&&\rho \PD{\bv}{t} + \rho (\bv \cdot \Grad) \bv =
-\Grad(P + \rho B^2/2) + \rho(\bB \cdot \Grad) \bB
+ \bB \Div(\rho\bB)\nn\\
&&\hspace{5em}
+ \lf[(\Gt-\f{1}{2}) \f{C}{8} +
\f{p(1-p)}{8 C} (C +3)\right] \nabla^2 (\rho\bv)
\nn\\
&&\hspace{5em}
+\left[(\Gt-\f{1}{2}) (\f{C}{4} + C_s^2) +
\f{p(1-p)}{4 C} (C -3)\right] \Grad[\Div(\rho\bv)] \nn \\
&&\hspace{5em}
+ \f{\sqrt{3}}{8} \f{p(1-p)(2p-1)}{C} \left[2 \Grad(\Div(\rho\bB)) -
\nabla^2(\rho\bB) \right], \lab{112}
\eeqy
and
\beqy
&&\PD{\bB}{t} + (\bv\cdot\Grad)\bB = (\bB\cdot\bf\Grad)\bv +
\f{\bv}{\rho}  \Div(\rho\bB)\nn\\
&&\hspace{5em}
+ \left[ (\Gt-\f{1}{2}) \f{3C}{8}
+ \f{3}{8}\f{p(1-p)}{C} (C -1) \right]\nabla^2(\rho\bB)\nn\\
&&\hspace{5em}
-\left[(\Gt-\f{1}{2}) \f{C}{4} +
\f{p(1-p)}{4 C} (C -3) \right] \Grad(\Div[\rho\bB)] \nn\\
&&\hspace{5em}
+ \f{\sqrt{3}}{8}\f{p(1-p)(2p-1)}{C}\left[ 2 \Grad(\Div(\rho\bv)) -
\nabla^2(\rho\bv) \right],   \lab{113}
\eeqy
where $P$ corresponds to the mechanical pressure.
We can readily notice that the unphysical terms
$\Grad(\Div(\rho\bB))$ and $\nabla^2(\rho\bB)$ in
(\ref{112}), and
$\Grad(\Div(\rho\bv))$ and $\nabla^2(\rho\bv)$ in
(\ref{113}) can be eliminated in the hydrodynamics limit
($p=0$), but more
importantly they can be removed for $p=1/2$, so that the MHD
macroscopic behavior can be maintained.

The extra terms in the macroscopic equations for $\rho$, $\bv$ and $\bB$
are also present in the $36$-bit model described earlier in this work.
The origin of these terms lies on the bidirectional streaming used
in these models. The most seriously
offending terms in (\ref{112}) and (\ref{113}) can be eliminated with the
symmetrically appealing choice of splitting the distribution in halves.
As mentioned in the previous section, the model should be symmetric
about $p=1/2$ and it is noticed that all
coefficients appearing in the macroscopic equations are indeed
invariant under the exchange $p \rightarrow 1-p$.

For this choice of $p=1/2$ (implying $C=3/2$) the values for the
transport coefficients are :
\beqy
\nu   &=& \f{3}{16} \Gt,   \lab{nu13}\\
\nu_b &=& \f{3\Gt}{2}\left(\f{1}{4} + \f{3}{12+\Ga}\right) -
\f{1}{4} \left( 1 + \f{9}{12+\Ga} \right), \\
\mu   &=& \f{9}{16} \left( \Gt -\f{4}{9} \right), \lab{mu13}\\
\mu_b &=& -\f{1}{8} \left( 3\Gt -2 \right), \lab{mub13}
\eeqy
where $\nu_b$ and $\mu_b$ are the bulk viscosity and the bulk resistivity,
respectively, and $\Ga$ is a free parameter introduced in the previous
section that is used to set the sound speed.
Notice that the ratio $\nu/\mu$ can be arbitrarily chosen by conveniently
adjusting the parameters $\Gt$ and $\Ga$.
By inspecting these expressions we can realize that the model,
unlike the hydrodynamics model, displays positive $\nu$ and $\mu$
beyond the threshold for stability ($\Gt=1/2$). Simple stability arguments
indicate that the parameter $\Gt$ should not be less than $1/2$, therefore
imposing a lower bound on the transport coefficients.

\section{Hartmann Flow}
We now turn to the application of the lattice Boltzmann
model we just described
to a linear magnetohydrodynamics problem, namely
Hartmann flow \cite{ferraroplumpton,MHD}. This problem represents
one of the  few MHD configurations that can be analytically
solved without the need of linearizations (the equations are linear),
with the additional assumption of constant density and
constant transport coefficients.

The Hartmann configuration
consists of a conducting liquid along a uniform channel,
in steady regime and under the action of a transverse
magnetic field. These flows can be used as
flowmeters, by measurement of the potential induced in the
fluids as it streams exposed to the external magnetic
field \cite{ferraroplumpton,shercliff}.
The fluid, assumed incompressible, is constrained to flow
horizontally in a very long, ideally infinite channel alongside
the x-direction. All relevant quantities, except the pressure,
are a function of only the transverse coordinate (to the
channel) $y$, $\bv=(v_x(y),0,0)$, $\bB=(B_x(y),B_0,0)$.
A uniform and time independent pressure gradient
is maintained along the channel direction to drive the fluid, so that
$p=p(x)$. The walls are located at $y=-L$ and $y=L$. Opposing to
the propelling pressure gradient is the viscosity of the fluid and the tension
in the magnetic field lines that resist the bending effect of the flow.

For this case, the incompressible MHD equations
\beqy
&& \PD{\bv}{t} + (\bv \cdot \Grad )\bv =
\f{1}{\rho}\Grad(p + \f{B^2}{2}) +
(\bB \cdot \Grad )\bB + \nu \nabla^2\bv, \nn \\
&&\PD{\bB}{t}  + (\bv \cdot \Grad )\bB =  (\bB \cdot \Grad )\bv
+ \mu \nabla^2\bB, \nn
\eeqy
can be reduced to the following linear system
\beqy
&&\nu \f{d^2 v_x}{dy^2} +
B_y \f{dB_x}{dy} - \f{1}{\rho} \f{dp}{dx} = 0,    \label{eq:5.19}\\
&&\mu \f{d^2 B_x}{dy^2} + B_y \f{dv_x}{dy} = 0,\label{eq:5.20}
\eeqy
where we are assuming that the system has reached a steady state;
the density $\rho$ is assumed uniform, and $B_y(\equiv B_0$ from now on)
corresponds to the known externally applied constant
magnetic field transverse to
the channel.
If non-slip (static plane walls) boundary conditions for the
velocity, and $B_x(-L)=B_x(L)=0$ for the magnetic field are applied,
the following analytic solutions can be found for the system
(\ref{eq:5.19}) and (\ref{eq:5.20}),
\beqy
v_x(y)&=&\f{f L}{B_0 \rho} \sqrt{\f{\mu}{\nu}} \coth(H)
\lf[ 1 - \f{\cosh(Hy/L)}{\cosh(H)} \r],   \label{hart-ux} \\
B_x(y)&=&\f{f L}{B_0 \rho}
\lf[\f{\sinh(Hy/L)}{\sinh(H)} -\f{y}{L}\r],  \label{hart-bx}
\eeqy
where the solutions depend on the dimensionless Hartmann number
$H\equiv B_0 L/\sqrt{\mu \nu}$, that measures the relative importance
of viscous and magnetic forces. In the above expressions $f= dp/dx$,
and represents the force driving the fluid down the channel.
It is easy to prove that in the limit of no external magnetic
field $B_0$ (that corresponds to $H=0$ if all other parameters are
maintained constant), the solution for $B_x$ is identically zero, and
$v_x = -(f L^2/2 \nu) [(y/L)^2 -1 ]$. This is the well-known solution
for a simple Poiseuille flow.

The boundary conditions imposed for the magnetic field imply that
the walls have a finite conductivity (they are neither perfectly conducting
nor perfectly insulating).
In this configuration (with these boundary conditions), the
Hartmann flow is operating as an electromagnetic flowmeter as we
will immediately show \cite{MHD}.

The current density $\bj$ can be
obtained from Ohm's law as $\bj = \Gs (\bE + \bv \times \bB)$,
where the conductivity $\Gs$ is the inverse of the resistivity $\mu$
in our units, and $\bE$ is the electric field. The only surviving
component of $\bj$ is the component normal to the plane of the flow,
and it can be computed as
\beq
j_z = \Gs (E_z + v_x B_0). \lab{j_z}
\eeq
On the other hand, from Maxwell's induction equation
we can get that $j_z(y)=-\p B_x/\p y$.
This expression for the current density can be immediately
integrated to obtain the total current across the channel $J_z$.
Noting that $B_x(y)$ is an odd function of $y$, or simply by
using (\ref{hart-bx}) we get
$$
J_z = -\int_{-L}^{L} \f{\p B_x}{\p y} dy = B_x(-L) - B_x(L) = 0.
$$
Combining with (\ref{j_z}) we obtain that
$$
\int_{-L}^L E_z = -B_0 \int_{-L}^L v_x(y) dy,
$$
from which it follows that
\beq
E_z=-B_0 v_M, \lab{E_z}
\eeq
where $v_M$ represents the mean velocity of the flow
across the channel. Consequently, $v_M$ can be computed
by measuring $E_z$. ($B_0$ is assumed to be known  since it is
externally applied.)

Now we turn to
the numerical
simulation of Hartmann flow, making use of the lattice
Boltzmann model with $12$ moving states described in
previous sections.
The simulation domain was for all cases a lattice of only $4$
cells in the $x$ direction of the problem $\times$ $60$ cells in the
transverse direction $y$. The $x$ direction coincides with the
direction ${\bf e}_1$ of the hexagons.
The system is initialized by setting the distribution functions
$\fas$ and $f_0$ to their equilibrium values given by a uniform density,
and a transverse magnetic field $B_y \neq 0$. $B_x$, $v_x$,
and $v_y$ are initially
zero.

The system is evolved by using the standard sequence of collisions and
streaming processes, with the addition of an intermediate step that acts
to generate an effective pressure gradient.
To achieve this, the distribution of moving particles at each cell
$\fas$ is redistributed so that the total density and the magnetic field
at the cell are unaltered, and the velocity receives a ``kick'' in the
direction of the flow. Care must be taken that the redistribution
of mass density along the different states does not push the
distribution function too far out of its equilibrium value, nor makes
it negative.
That is, for the problem under consideration, a forcing function
must be constructed that tries to increase $\rho v_x$ while
{\em not} changing $\rho v_y$, $\rho \bB$, or $\rho$.
To explicate the dynamics of this procedure, let us recall the definitions
of the macroscopic quantities,
\beqy
\rho &=& f_0 + \sum \fas \nn\\
\rho \bv &=& \sum [(1-p)\ea + p\es] \fas  \nn\\
\rho \bB &=& \sum [r\ea + q\es] \fas.   \nn
\eeqy
Therefore, if at a certain microscopic time $T$ the distribution
function at a specific cell is $\fas$ before this ``forcing'' scheme, and
$f_a^{\Gs\prime}=\fas + F^\ast C_{a\Gs}$ after the distribution is kicked, we
wish to find a quantity $F^\ast C_{a\Gs}$ with the constraints
\beqy
\rho = \sum_{a\Gs} f_a^{\Gs\prime} &=& \sum_{a\Gs} \fas + \GD\rho\\
\sum_{a\Gs} f_a^{\Gs\prime} [(1-p)\ea + p\es]&=& \sum_{a\Gs} \fas[(1-p)\ea +
p\es] +
\rho (\GD\bv) \\
\sum_{a\Gs} f_a^{\Gs\prime}[(r\ea + q\es] &=& \sum_{a\Gs}
\fas[(r\ea + q\es] + \GD(\rho\bB),
\eeqy
where $\GD$ represents a ``small'' change.
As stated above, we are seeking $\GD(\rho)=0$, $\GD(\rho\bB)={\bf 0}$.
Using that $f_a^{\Gs\prime}=\fas + F^\ast C_{a\Gs}$ and adding over $\Gs$,
Eqs. 51-53 become
\beqy
\GD(\rho) &=& 0 = \sum_{a} (C_a^1 + C_a^2) =0,\label{aa}\\
\GD(\rho \bv) &=&  F^\ast (1-p) \sum_{a} (C_a^1 + C_a^2) \ea +
                F^\ast p \sum_{a} (C_{a-1}^1 + C_{a+1}^2) \es,
\eeqy
and
\beq
\GD(\rho \bB) = 0 = r \sum_{a} (C_a^1 + C_a^2) \ea +
               q \sum_{a} (C_{a-1}^1 + C_{a+1}^2) \es.    \label{aa1}
\eeq
Combining (\ref{aa}) and (\ref{aa1}) we obtain
\beq
\GD(\rho \bv) = (1-p-p\f{r}{q}) \sum_{a} (C_a^1 + C_a^2) \ea.
\eeq
Choosing $\GD(\rho v_x)=1$ and $\GD(\rho v_y)=0$
the following set of equations can be found
\beqy
& &\sum_{a} (C_a^1 + C_a^2) = 0 \nn \\
& &C_2^1 + C_2^2 + C_3^1 + C_3^2 - C_5^1 - C_5^2 - C_6^1 - C_6^2 = 0 \nn\\
& & (1-p-p\f{r}{q})\lf[ C_1^1 + C_1^2 + \f{C_2^1}{2} + \f{C_2^2}{2} \r.\nn\\
& & \hspace*{8em}- \f{C_3^1}{2} - \f{C_3^2}{2} - C_4^1 - C_4^2 - \f{C_5^1}{2}
- \f{C_5^2}{2} + \f{C_6^1}{2} +\lf. \f{C_6^2}{2}\r] = 1,
\eeqy
for which a {\em particular} solution is
\beqy
C_1^1 &=& C_1^2 = -C_4^1 = -C_4^2 = 1 \\
C_5^1 &=& -C_6^2 = \f{3-p}{2-p} \\
C_5^2 &=& -C_6^1 = \f{3-2p}{2-p},
\eeqy
where we used that $r/q=-(1+p)/(2-p)$.
By appropriately choosing $F^\ast$ we are able to control the strength
of the forcing scheme. We note that the pressure gradient produced in this
way is uniform, and that it follows the spirit of the CA interpretation of
cell populations as ``particles.''
When this process is put into action, the total momentum in the
direction that is being forced is seen to increase, until the
driving force is balanced by the action of the viscosity of the fluid
and the reluctance of the magnetic field lines to be bent, and the system
reaches a stationary regime.

The boundary conditions were implemented by setting
$v_x$, $v_y$ and $B_x$ equal to zero in the first and last layers (i.e.,
$y=0$ and $y=L$),
combined with a periodic streaming. An alternative way to achieve
the boundary conditions would be to cancel the periodic streaming
of the populations, and to make the ``particles'' bounce off
the walls reversing the velocity at those cells while keeping
unchanged the magnetic field.

Simulations of the Hartmann flow were carried out for Hartmann number
$H$ $=0$ (zero magnetic field case), $1,3,5,8$, and $13$.
The results are presented in Fig. 1
and Fig. 2.
 Figure 1
displays all the velocity profiles $v_x(y)$ plotted versus $y$,
for all six values of $H$, and compared with the analytical
solution (\ref{hart-ux}). A very good agreement between the
latter and the computational results is obtained for this range of
values of H.

The flattening of the velocity profile can be understood
in several ways \cite{ferraroplumpton,MHD,shercliff}.
{}From the linearized MHD equations (\ref{eq:5.19})
and (\ref{eq:5.20}), it can be seen that the transverse magnetic field
tries to eliminate vorticity. If the magnetic forces dominate (i.e.,
for large Hartmann number $H$), then the velocity profile tends to
flatten.\cite{footnote2}
However, the velocity profile cannot be flat all across the channel,
because the velocity at the walls vanishes, so there must be a region
within a certain distance from the walls on which the gradients of the
velocity are very large, i.e., the region where the vorticity
(produced by the boundaries) is confined to.

Alternatively, the flattening of the velocity profile could be
understood as follows. Combining (\ref{j_z}) and (\ref{E_z}) we obtain
$$
j_z = \Gs B_0 (v_x - v_M).
$$
The magnetic force on the fluid is given by the Lorentz force
$\bF_L=\bJ\times\bB$, that in our case reduces to
$(F_L)_x = -\Gs B_0^2 (v_x(y) - v_M)$, and therefore the flow
tends to slow down where $v_x > v_M$ and tries to speed up
where $v_x<v_M$, producing a flattening of the velocity profile.

The high degree of agreement seen for the velocity profile
across the channel
is also observed for Figure 2,
 that shows the
same comparison for $B_x(y)$, for $H=1,3$ and $13$
only. For this figure, only results for three different Hartmann
numbers were included for the sake of clarity, since the effect of
$H$ on $B_x$, is not as marked as it is for $v_x$. We note that the
fit for the cases not shown is as good as those displayed in
the figure. There is a point we would like to stress about the material
presented in this figures. These are not ``fits'' in the usual sense.
The Hartmann number $H$ is here constructed from $\nu$ and
$\mu$ which come from the Chapman-Enskog theory.
The analytic solutions (\ref{hart-ux}) and (\ref{hart-bx})
depend only upon $f$, $\mu$, $\nu$, $B_0$, $L$, and $\rho$.
The forcing $f$ is fixed; $B_0$ is fixed initially as are $\rho$ and
the simulation size $L$. With this subtlety in mind, the
solutions have {\em no} free parameters to adjust.

Some caution must be exerted when calculating the Hartmann number
corresponding to the simulation, since the exponential character
of the solutions would make them very sensitive to a small departure
from
the actual value of $H$. In particular, the width of the channel $2L$
should be evaluated as $(N_y -1) \sqrt{3}/2$, taking into account
the $x-y$ ratio for the hexagons, and the position of the boundary
in the simulation is at $y=1$ instead of $y=0$.


The Hartmann number $H=B_0L/\sqrt{\mu \nu}$ was varied by changing
the strength of the magnetic field $B_0$. The size ($60$ cells)
was kept constant for convenience in the manipulation of data.
The kinematic viscosity $\nu$ and resistivity $\mu$ can be obtained
as a function of the relaxation time $\tau$. From
Section III we
recover the expressions $\nu = 3\tau/16$ and
$\mu = 9\tau/16 - 1/4$.
Only $\tau =1$ was used for all the simulations in this section,
therefore the system is forced to equilibrium in each iteration.
The ``forcing''
coefficient was set to $F^\ast=2 \times10^{-5}$, thus being
sufficiently small (comparing $F^\ast C_a^\Gs$ with the mean value of the
density,
$\rho_0 = 3.9$) to ensure that the distribution function $\fas$
will be only slightly departed from equilibrium during the
``forcing'' step.

The limitations for the range of $H$ that the model will be able
to accurately reproduce are as follows. 1) for the upper bound,
we can see from (\ref{eq:5.19}) and (\ref{eq:5.20}) that we can
estimate the width of the
boundary layers, in which the Lorentz force is comparable to the viscous
forces, as $\Gd\sim \sqrt{\nu\mu}/B_0$, from which we gather that
$\Gd/L \sim H^{-1}$. When $H\gg1$,  the thickness
$\Gd$ becomes very small (the region
of nonzero velocity gradients is confined to a very narrow layer
away from the walls), and chances are that we need to increase
the width of the domain to resolve $\Gd$. For our simulations, for which
$60$ cells across the channel were used for all cases, we observed that
for $H>30$ the boundary layer thickness $\Gd$ is of the order
of one cell. Therefore, if simulations with higher values for $H$
are desired, $L$ should be increased. We note that this is not a
limitation of the lattice Boltzmann simulation scheme, but rather a
resolution constraint: the computation of the analytical solution
presents the same flaws. 2) for the lower bound, the limitation is
given by the roundoff error of the machine since for $H\ll 1$ we are
forced to use very weak external fields $B_0$.

The one dimensional nature of the problem is highlighted by the fact
that the exact behavior of the solutions was reproduced for a domain
that was $4$ cells long in the fluid direction $x$. As a matter of fact,
we observed that the same results can be obtained with a length
of only one cell, making the domain truly one-dimensional, and
supporting the hypothesis made for the derivation of the lattice
Boltzmann approach, that the population of the cells correspond to
ensemble averages of the discrete populations used for the
Cellular Automaton approach.

\section{2D Magnetic Reconnection}
In the previous section, we argue in favor of our lattice Boltzmann
MHD model, by comparing its solutions for the Hartmann flow problem,
 with the analytically obtained solutions.
The results are encouraging to the extent that is very
difficult to observe with the naked eye in Figures 1 and 2
departures of the LBE solution
from the theoretical ones. This was the case for a wide range
of values of the Hartmann number, the only
parameter in the problem.
Optimistic as we might be, we recognize that
Hartmann flow
is essentially a {\em one-dimensional} and {\em linear}
problem. Thus, it is our intention in this section, to test the validity
of the $13$-bit LBE model for a situation that is both {\em two-dimensional}
and {\em nonlinear}. The configuration we chose is the 2D MHD
``sheet pinch''.
Before presenting results for the present 13-bit model,
we recall that the reconnection configuration
had been chosen previously as a test problem for
$36$-bit MHD model
of Ref. \cite{ccmm}.
The authors found that the LBE solutions were qualitatively
correct, and they showed features of the evolution associated
with nonlinear effects of the sheet pinch dynamics. However, no claim
of comparison with the ``real'' solutions, or solutions from other
numerical methods was made.

The ``sheet pinch'' is a magnetohydrodynamic configuration characterized
by an inhomogeneous magnetic field that changes markedly in a very narrow
region, thus producing very strong currents.
This arrangement can be encountered in a variety of important physical
phenomena as solar flares, and the earth's magnetic field, reversing its
sign embedded in the solar wind.
It is not the goal of this section to discuss in detail the physical
processes in a magnetofluid undergoing a reconnection process, a vast
literature is available. Central are the theoretical efforts of
Dungey \cite{dungey}, Parker \cite{parker}, and Sweet \cite{sweet}.
For a discussion of the reconnection process, including the role played
by the fluctuations from the point of view of turbulence, see Matthaeus and
Montgomery \cite{matth-montg1981}, and Matthaeus and
Lamkin \cite{matth-lamkin1986}.

Our approach to the ``sheet pinch'' problem will be
similar to the
procedure followed in another recent test of the LBE method
\cite{dom}, in which LBE and spectral method solutions for a 2D
hydrodynamic shear layer were compared in detail.
Here, we will briefly describe the reconnection
runs from a technical point of view,
and then we will move on to present and contrast the results obtained
with both LBE and spectral methods.
\subsection{The Sheet Pinch Simulations}
The idealized sheet pinch consists of a uniform magnetic field reversing
 sign in a very thin zone, much in the same way as the velocity
swaps its direction in the idealized nonlinear
Kelvin-Helmholtz instability.
This configuration gives rise to a current sheet because
$j = (\Curl \bB)_z$, with $j$ the
current density in the $z$ direction. Therefore, the initialization
of the sheet pinch was done,
in a $2\pi\times 2\pi$
simulation box, with a spectral truncated
representation of delta functions located at
$y=\pi/2$ and $y=3 \pi/2$ (with opposite signs)
for the current, including wavevectors
$k=1$ through $31$.

An uninteresting evolution follows unless some non-sheet pinch modes
are excited, to initiate the nonlinear couplings. The current
density and the vorticity Fourier modes with $1 \leq k \leq 15$ where
excited with random phases and with spectra of $k^{-3}$ for
high $k$ for both, kinetic and magnetic energy. The ``noise'' spectrum was
peaked at $k=3$ for both kinetic and magnetic energy, and added about
$1\%$ of the energy already present in the ideal sheet pinch.

For the spectral run the $z$ components of the vorticity
${\bf \Go} = (0,0,\Go)$ and the vector potential $\ba = (0,0,a)$, are
evolved according to the equations
\beqy
\PD{\Go}{t} &+& \bv \cdot \Grad \Go = \bB \cdot \Grad j + \nu \Grad^2 \Go \\
  \PD{a}{t} &+& \bv \cdot \Grad a = \mu \Grad^2 a,
\eeqy
where $\mu$ is the resistivity and $\nu$ is the kinematic viscosity.
The fields $\bv$ and $\bB$ are in the plane $x,y$ depending solely
on those coordinates. The scalar functions $\Go$ and $a$ are related
to $\bv$ and $\bB$ by
$\Go \bz = \Curl \bv$, and $\bB=\Curl a \bz = \Grad a \times \bz$,
where $\bz$ is the unit vector in the direction normal to the
$x,y$ plane.
The current density is $\Grad^2 a = -j$, and the vorticity is related
to the stream function by $\Grad^2 \psi = -\Go$.
The spectral run is of the Fourier-Galerkin type
and has a resolution of $128 \times 128$.

For the LBE run we have to specify the initial density, velocity
field, and magnetic field. The velocity field is obtained from the
relation $\bv=\Grad\psi\times\bz$, the stream function $\psi$ being obtained
from the solution to $\Grad^2 \psi = -\Go$, with $\Go$ being the initial
vorticity used in the spectral run.
Similarly, to initialize $\bB$, we numerically evaluate
$\bB= \Grad a \times \bz$ in Fourier space, $a$ being the initial
vector potential used for the spectral run. The initial density is set to
a uniform value $\rho = \rho_0 = 3.9$. The initial fields are used to
evaluate the equilibrium distribution function $\fas$ for this model,
given by Eqs. (\ref{faseq}) and (\ref{f0seq}),
for these initially specified fields.
{}From then on, the LBE sequence of streaming and collisions
is performed to evolve
$\fas$.

A short digression is needed at this point to justify the choice
of $\nu$ and $\mu$ for the spectral run (the inverse of the Reynolds
number $R$, and the magnetic Reynolds number $R_m$, respectively,
in the set of units we are using), and
the relaxation parameter $\tau$ and the simulation size, for the
LBE simulation.
Since it is our goal
to test the MHD LBE model in a situation that
involves strong nonlinear interactions,
it is desirable to perform a simulation with
the Reynolds numbers $R$ and $R_m$ as large as possible.
The limitation is imposed by the LBE scheme. In Section III,
explicit expression for the resistivity and viscosity of the LBE model
were found:
\beqy
\nu_{LBE} &=& \f{3}{16} \Gt \\
\mu_{LBE} &=& \f{9}{16} \Gt -\f{1}{4}.
\eeqy
For stability reasons, the parameter $\tau$ is
constrained to be
$\tau > 1/2$. On the other hand, for attaining small $\nu_{LBE}$ and
$\mu_{LBE}$, and thus high $R$ and $R_m$, the relaxation parameter $\tau$
should be as small as possible. Choosing $\tau$ slightly larger than $1/2$,
to ensure stability, the values of $R$ and $R_m$ will be dictated
essentially, by the simulation size. To strike a compromise between
a computationally reasonable simulation size,
and the degree of nonlinear activity, we use
a $512 \times 512$ resolution domain, for the LBE simulation.
The characteristic speed in our problem is the Alfv\'en speed,
coinciding with $\sqrt{\la B^2\ra}$ in our units.
Consequently, the magnetic Reynolds number is given by
\beq
R_m = \f{B L}{\mu} = 0.1 \times \f{512}{2\pi} \times \f{1}{\mu},
\eeq
where $B=\sqrt{\la B^2\ra}=0.1$ initially.
Similarly, for the mechanical Reynolds number we obtain,
\beq
R = \f{B L}{\nu} = 0.1 \times \f{512}{2\pi} \times \f{1}{\nu}.
\eeq
For $\tau = 0.5001$ we get $R= 86.9$ and $R_m = 260.3$. The reciprocal
of these numbers are those used for $\nu$ and $\mu$, respectively, for the
spectral run.
The factors of $2\pi$ are necessary because the length $L$ used to define
$R$ and $R_m$ for the spectral simulation is the physical
length of the simulation box, divided by $2\pi$, i.e.,
$L=2\pi/2\pi=1$. Thus, for the LBE simulation simulation
we need to use $512$ (the physical size) divided by $2\pi$.
Note that for the LBE run, because we are using the hexagonal lattice,
the typical length used for the above evaluations, is somewhat ambiguous
due to the $\sqrt{3}/2$ ratio between the $y$ and $x$ directions.
Moreover, the system represented by both methods are
{\em physically} different.

Last, before turning to the comparison of the results obtained for the
two runs, and to be able to relate physical processes
observed in both simulations, we need to find a relationship between the
spectral and LBE characteristic times. Let us write,
\beq
\f{T_{LBE}}{T_{SP}} = \f{L_{LBE}/L_{SP}}{U_{LBE}/U_{SP}} =
                                        \f{512}{2 \pi} \f{1}{0.1},
\eeq
where $U_{LBE}=0.1$ is the characteristic speed of the LBE model, given
by the Alfv\'en speed, and similarly $U_{SP}=1$ for the spectral method
run, that tells us that $T_{LBE} = 814.87 \; \;T_{SP}$.

The runs were carried out up to about ten spectral-method
characteristic times.
Periodic boundary conditions were imposed for the simulations.
\subsection{Comparison and Discussion of the Sheet Pinch Runs}
The evolution of the sheet pinch dynamics, being an MHD system,
is much more complex than its hydrodynamics counterpart.
There are more dynamical variables evolving coupled to each other, and
there is a larger parameter space. For example,
the relative values of $R$ and $R_m$
may influence the dynamics.
Nevertheless, there are some global features thought to
be similar for
most decaying, 2D, incompressible MHD flows in periodic geometry.
For example, there are three rugged
invariants \cite{fyfemontg}, that is constants of the nondissipative
evolution that survive the truncation
in $k$ space (Galerkin approximation). Those are \cite{fyfemontg},
the total energy $\la E\ra= \sum_\bk [\Go^2(\bk)/k^2 + a^2(\bk) k^2]$,
the cross helicity $H_c=\la \bv\cdot\bB\ra=\sum_\bk \Go(\bk) \,a(\bk)$,
and the mean square vector potential $A=\la a^2 \ra=\sum_\bk a^2(\bk)$,
where $\la\cdots\ra$ denotes a volume average.

In Fig. 3
we present time histories of bulk quantities that characterize the
turbulence for both the
spectral method and the LBE run. The continuous line corresponds
to the spectral method simulation. Panels a) and b)  of
Fig. 3
 display the
 evolution of  $A$ and $E$, respectively. Both quantities,
which are
ideally conserved, decay monotonically due to the presence
of nonzero
dissipation coefficients $\mu$ and $\nu$. The slow decay of
$A$ suggests
that there is a dynamic redistribution of this quantity in favor
of larger scales. In Fig. 3c)
 we present the
evolution of the kinetic energy. The kinetic energy $E_k$,
initially $1\%$ of the total energy, decreases throughout the
run, displaying
some small bursts of activity that would be more pronounced
for larger
$R$ and $R_m$ \cite{matth-lamkin1986}.

In panels d) and e) of Fig. 3
, we show the evolution
of the enstrophy $\GO$ and the mean square
current $J=\la j^2 \ra$. These two quantities highlight the activity
in small spatial scales, and for these runs they rapidly decay
due to the relatively high viscosity and ohmic dissipations.
Again, both quantities present a more ``bursty'' shape for more
turbulent systems \cite{matth-lamkin1986}.

We now turn our attention to comparison of contour plots.
In Fig. 4
 we exhibit plots of constant magnetic field
lines (constant $a$),
for times approximately equal to one, seven, and ten.
Emergence and subsequently growth of magnetic islands can
be seen.
For the same times, in Fig. 5
, contours of
constant vorticity are displayed. Although the
distribution of vorticity consists of small, nonlocalized structures,
it can be seen that this quantity rapidly organizes in the region
of the $X$-points to form quadrupole-like structures. The vorticity
plots, combined with the stream function contour plots, shown in Fig. 6
, provide a
consistent picture of part of the activity of the magnetofluid
in the reconnection zone. Jets of fluid are seen to come into the
``hot'' area (higher speeds are represented by denser $\psi$ lines)
from the strong-field sides of the $X$ (up and down in
our plots), and out through weak-field corners (sides of the $X$).
These jets are responsible for the bursts that suggest themselves
in the evolution of the kinetic energy.
This is essentially a pressure-driven effect due to the steep gradients
of the magnetic field present near the neutral sheet; the fluid
finds itself pushed by the magnetic pressure into the neutral zone, and,
being incompressible, it has no ``choice'' but to turn into the weak-field
region producing the four vortices seen in the plots, at the
$X$ points.

The last set of plots from the spectral and LBE runs
(Fig. 7), shows
contours of constant current density. Initially the current is
concentrated, more or less uniformly, along both current sheets. As the
system evolves, and the magnetic field lines start to reconnect, we see
a tendency for filaments of current density to form. The regions with
current filaments will participate in a significant part of the energy
dissipation due to finite resistivity.

We conclude from the examination of these plots, that this hexagonal,
two-dimensional MHD LBE model, is capturing the basic
mechanisms of MHD dynamics. From the contour plots, we see that
once a structure of one of the quantities is identified in one of
the plots corresponding to the spectral run, a very similar
structure can be also seen in the corresponding LBE plot.
Similarly, the LBE tracks the evolution of relevant bulk quantities
very closely. We would like to point out that a {\em perfect}
agreement is {\em not}, in fact expected. Additional refinements
could be introduced in the LBE simulation to still further reduce the
gap between the results from both simulations.
For example, it was mentioned above that the spectral simulation
box, and the LBE domain are physically different systems, due to the
effects coming from the hexagonal lattice.
This will certainly affect the magnitude of the mechanical and magnetic
Reynolds numbers. There are, at least, two alternative ways of
getting around this difficulty. One of them would be to make use
of the similar MHD model derived based on the square lattice,
which is described in Appendix B.
(The earlier comparison of the LBE and spectral method
hydrodynamic shear layer dynamics \cite{dom} employed
an LBE on a square lattice.)
Although the hexagonal LBE requires less
memory, this is not a decisive advantage.
Another possibility would be to abandon the use of the same number
of cells in both dimensions ($N_x=N_y$), and to choose, for example
$N_y = N_x \sqrt{3}/2$. This choice would have introduced
complications in diagnostics currently based upon
Fast Fourier Transforms.
Instead, considerable larger amounts of data would
be required to be kept for later analysis.

An examination of the divergence of the magnetic field is mandatory,
to make sure that monopoles are not being created by the model,
thus casting doubt on the results. To this end, we decompose
in Fourier space the magnetic field $\bB(\bk)$ in its
longitudinal component $B_L$, and its transverse component
$B_\perp$.
A similar examination was carried out for the velocity field in the
shear layer LBE study \cite{dom},
as a way to quantitatively measure the
compressibility of the flow. We calculate $B_L$ and $B_\perp$ as
\beqy
B_L^2     &=& \sum_\bk \f{|\bk \cdot \bB(\bk)|^2}{k^2}, \\
B_\perp^2 &=& \sum_\bk \f{|\bk \times \bB(\bk)|^2}{k^2}.
\eeqy
This ratio is a good measure of the amount of ``monopolar''
(longitudinal) activity as compared with the transverse component,
containing most of the energy. In Fig. 8
 we show the evolution of
the ratio $B_L/B_\perp$. We readily notice that the overall
tendency of this quantity is to decrease, and that by the
end of  the run the ``amount'' of $B_L$ is about one part in
one thousand.

The nonzero initial value of $B_L$
is attributed to the non-square nature of our LBE simulation box.
Exactly the same field that produces $\Div \bB=0$ for the square spectral
simulation run, produces nonzero divergence
on the hexagonal lattice.
What is encouraging about this picture is that the LBE dynamics seems to
possess self-adjusting mechanisms that reduce the amount of monopoles,
much
in the same way as Chen \etal \cite{hudong-bill2} discuss in
the context of their $36$-bit MHD CA model.

Finally, we turn to a brief discussion of the efficiency of the method.
At the end of the previous section,
we noted that although the transport coefficients are directly
controllable via the relaxation parameter $\tau$, the threshold of
stability with respect to $\tau$, is higher than the value of
$\tau$ needed to make $\nu$ and $\mu$ zero. This technical problem
limits the Reynolds numbers that can be attained in the MHD LBE for
fixed grid size.
In the
case of the hydrodynamic shear layer LBE \cite{dom},
the simulation domain
was chosen with the objective of resolving the spatial structure of the
turbulent activity, much in the same way as it is done for a spectral
method simulation. Thus, at the same Reynolds number,
an LBE and spectral method
hydrodynamics simulation can have about the same grid size.
For the MHD LBE, the size was also determined by the requirement of
matching the Reynolds numbers with the spectral code.
This $512 \times 512$ LBE simulation was run on a CRAY-YMP computer, in
the San Diego Supercomputer Center, and needed about
$12$ minutes per characteristic time (about $800$
LBE microscopic times). The spectral run, $16$ times smaller ($128
\times 128$), took
about ten times less CPU time.
Nevertheless, it should be noted that the numerical efficiency
of LBE-type computations is greatly enhanced
in massively parallel computers, due to its local dynamics.

\section{Discussion and Conclusions}
In this paper we have introduced a model for
simulation of 2D MHD with the lattice Boltzmann equation technique.
The idea of propagating the distribution at a given state
into two directions associated with the velocity and magnetic fields,
had been previously used for obtaining 2D MHD using CA dynamics
\cite{hudong-bill,hudong-bill2}, and later extended to LBE
\cite{ccmm}. In the present scheme, by utilization of the same
idea combined with the flexibility of the LBE scheme a
significantly more efficient and simpler method is obtained for simulation of
2D MHD. The improvement is two-fold; first, the number
of discrete velocities is reduced in our model from 37 to 13, only.
Second, the algorithm for the evolution of the present model is simplified
by requiring a ``forward'' streaming, as explained in the text.
These models possess the same microscopic symmetries necessary for
guaranteeing the correct long wavelength, low frequency behavior.
The theory for the model was presented, including
the Chapman-Enskog expansion procedure to obtain second-order effects.
In passing, we note that the simplicity of the model is apparent
when evaluating all the second-order contributions displayed in Section III.

Evidence of correct MHD behavior was introduced in Sections IV and V, where
the model is applied to reproduce a linear problem (steady Hartmann flow), and
a nonlinear problem (evolution of the 2D sheet-pinch).
For the former, the performance of our model is extremely good, for a
reasonably wide range of the Hartmann number, that parameterizes the problem.
The second numerical test of the model behaves reasonably well.
Nevertheless, it should be mentioned that this more stringent test
exposes what might be the most serious deficiency of the model, namely
the inability to achieve relatively low transport coefficients.
This disadvantage hurts the potential use of this model for
highly turbulent simulations. Clearly further investigation into this
matter is required. At the moment we ignore the fundamental reason of
this crossover
(in terms of the relaxation time $\Gt$) between the stability threshold
($\Gt > 1/2$), and the region of low values of viscosity and resistivity that
occurs for $\Gt < 1/2$ (see Eqs. (\ref{nu13}) and (\ref{mu13})),
and whether this effect is induced by the
special streaming used for the model, or the deficiency could be cured
by choosing a more flexible collision operator.
The LBE model for hydrodynamics
(with single-time-relaxation collision operator) does not share
this inconvenient feature since
$\Gt > 1/2$ is both a condition for numerical stability and for
positivity of the viscosity. Another two desirable properties that an
improved MHD LBE model could have are as follows; first, it would be
convenient to have independent control on the viscosity and
resistivity (in the present model, the choice of the relaxation
time $\Gt$ determines both $\mu$ and $\nu$). Second, and more important,
in the present model, and in other MHD models we made reference to in
this report, the divergenceless property of the magnetic field is not imposed.
It is seen, however (see Ref \cite{hudong-bill2}) that the model
possesses some self-adjusting mechanisms that diffuse away the solenoidal
component of the magnetic field, as displayed in Section V,
for the sheet-pinch simulation.

In spite of these improvable aspects, we believe that the 13-speed model
is capturing the essential features of the equations for MHD
with a minimum number of degrees of freedom, and that it is a valuable
tool for non-turbulent regimes. Moreover, although the model was formulated
explicitly for 2D in the present paper, its extension to 3D with a reasonably
low number of degrees of freedom does not seem to pose any serious difficulty.

\section*{Acknowledgements}
Useful discussions with Dr. H. Chen are gratefully acknowledged.
This research has been supported by National
Aeronautic and Space Administration Innovative
Research Program Grant NAGW-1648, and by the U.S. Department of
Energy at Los Alamos National Laboratory.
Computations were supported by the National Science Foundation
at the San Diego Supercomputing Center.

\section*{Appendix A: Useful Tensorial Relationships}
The relevant tensorial relationships used extensively for the
 derivation of the fluxes of density, momentum, and magnetic field
in Section II are:
\begin{eqnarray}
& &\sum_{a,\Gs} (\ea)_i (\ea)_j (\ea)_k (\ea)_l =
\sum_{a,\Gs} (\es)_i (\es)_j (\es)_k (\es)_l =
2 \times \f{3}{4} \Delta_{ijkl}  \nn \\
& &\sum_{a,\Gs} (\ea)_i (\ea)_j (\ea)_k (\es)_l =
\sum_{a,\Gs}(\es)_i (\es)_j (\es)_k (\ea)_l =\f{3}{4}\Delta_{ijkl} \nn \\
& &\sum_{a,\Gs} (\ea)_i (\es)_j (\ea)_k (\es)_l =
-\f{3}{4} \Delta_{ijkl} + \f{9}{2} \delta_{ij} \delta_{kl} \nn \\
& &\sum_{a,\Gs} (\ea)_i (\ea)_j (\ea)_k =
\sum_{a,\Gs} (\es)_i (\es)_j (\es)_k = 0                 \lab{tensors} \\
& &\sum_{a,\Gs} (\ea)_i (\ea)_j = \sum_{a,\Gs}(\es)_i(\es)_j =
6 \delta_{ij} \nn \\
& &\sum_{a,\Gs} (\ea)_i (\es)_j = 3 \delta_{ij} \nn \\
& &\sum_{a,\Gs} (\ea)_i = \sum_{a,\Gs} (\es)_j = 0 \nn,
\end{eqnarray}
where $\delta_{ij}$ is the Kronecker delta, and
$\Delta_{ijkl} = \delta_{ij} \delta_{kl} +
\delta_{ik} \delta_{jl} + \delta_{il} \delta_{jk}$.

\section*{Appendix B: Simulating 2D MHD on the ``Square'' Lattice}
Moving from modeling hydrodynamics to modeling MHD
requires the inclusion of new force terms, and more importantly a new
whole equation to follow the evolution of the magnetic field. This equation
involves terms like $(\bv\cdot\Grad) \bB - (\bB\cdot\Grad)\bv$, and
the technique that has worked in the $36$-bit and in the $13$-bit
models, has been splitting the distribution in different directions during
the streaming part of the evolution. This parting of the distribution produces
a mixture of directions that allows to include terms mixing $\bv$ and $\bB$
and with different signs. Once this fact is recognized, it
is not hard to
generalize the same idea to the square lattice.

In this section we would like to simply display a method
for modeling 2D MHD on the square lattice. This would have a slight
advantage and a slight disadvantage over the hexagonal lattice
($13$-bit model). The latter is that the ``square'' model requires
more memory than the hexagonal model: we need to keep track of $13$
real numbers per cell, for the hexagons, whereas for the squares the
requirements increase to $17$ real numbers per cell.
On the other hand, the square lattice provides a ``better'' simulation
domain, in a geometrical sense, for some systems, unlike the hexagonal
case, for which the basic cell has different physical lengths in the two
directions.

We now turn to a brief description of this alternative approach.
If one of the square cells is located at $\bx$, its nearest
neighbors are located at the face-centers ${\bf x} + {\bf c}^I_a$,
for $a = 1,2,3,4$, with ${\bf c}^I_a\equiv (\cos{(a-1)\pi/2},
\sin{(a-1)\pi/2})$,
and the vertices of the square centered about $\bf x$, i.e.,
${\bf x} + {\bf c}^{II}_a$, for $a = 1,2,3,4$, with
${\bf c}^{II}_a\equiv \sqrt{2}(\cos{(a-1/2)\pi/2}, \sin{(a-1/2)\pi/2})$.

There will be three distribution functions: one that streams in the
lattice indicated by the superscript $I$, a second one that is moved in the
lattice $II$, and a ``stopped'' distribution.
Therefore, the streaming part of the evolution can be
represented by the expression,
\beq
\fab^{K}(\bx,T) \rightarrow \f{1}{2} \fab^{K}(\bx+{\bf c}_a^K,T+1)+
                   \f{1}{2} \fab^{K}(\bx+{\bf c}_b^K,T+1),
\eeq
where $K=I$ or $II$, $a=1,2,3,4$, $b=a+1$ or $a-1$ (modulo 4),
and $T$ is the discrete lattice time.

The macroscopic quantities we are interested in following, the
density, velocity, and magnetic field, are defined below,
\beqy
\rho &=& f_0 + \sum_{a,b,K} \fab^K \\
\rho \bv&=&\f{1}{2}\sum_{a,b,K} ({\bf c}_a^K + {\bf c}_b^K) \fab^K \\
\rho \bB&=&\sum_{a,b,K} q_K (-{\bf c}_a^K + {\bf c}_b^K) \fab^K,
\eeqy
where $q_1=1/2$, and $q_2=1$ and $f_0$ represents the stopped
distribution.

During the collisional part of the evolution is when the three types of
distributions ``see'' each other. We use the single time relaxation
approximation with parameter $\tau$, so that the discrete
kinetic equation obeyed by the system is,
\beqy
\fab^K(\bx,T) = \f{1}{2} [\fab^K (\bx -{\bf c}_a^K,T-1) +
\GO_{ab}(\bx -{\bf c}_a^K ,T-1) ] + \\ \nonumber
\f{1}{2} [\fab^K (\bx -{\bf c}_b^K ,T-1) + \GO_{ab}(\bx -{\bf c}_b^K,T-1)],
\eeqy
where
\beq
\GO_{ab} = -(f_{ab}^K -f_{ab}^{K (eq)})/ \tau.
\eeq
The procedure to get the macroscopic MHD equations is familiar to us
at this point. The only thing that is left to complete the definition
of the model is to specify the distribution functions,
\beqy
\fab^K&=&\rho d_K \lf[1+ \f{1}{24d_2} ({\bf c}_a^K+{\bf c}_b^K)\cdot \bv +
 \f{\Ga_K}{24d_2} ({\bf c}_a^K+{\bf c}_b^K)\cdot \bB +
 \f{1}{32d_2} ({\bf c}_a^K\cdot \bv)^2 \r.  \nn   \\
&&+ \f{1}{32d_2} ({\bf c}_b^K\cdot \bv)^2 +
\f{1}{24d_2} ({\bf c}_a^K\cdot \bv)({\bf c}_b^K\cdot \bB) -
\f{1}{24d_2} ({\bf c}_a^K\cdot \bB)({\bf c}_b^K\cdot \bv) \nn \\
&&\hspace{4em}\lf.+ \f{1}{8d_2} ({\bf c}_a^K\cdot \bv)({\bf c}_b^K\cdot \bv)-
\f{1}{8d_2} ({\bf c}_a^K\cdot \bB)({\bf c}_b^K\cdot \bB) \r]  \\
\fab^0&=&\rho d_0 \lf[ 1-\f{3}{2d_0}v^2\r],
\eeqy
where
\beqy
d_0   &=& 1-40d_2    \hspace{6em}d_1=4 d_2 \nn \\
\Ga_1&=& 1            \hspace{9em} \Ga_2 = \f{1}{2}, \nn
\eeqy
$K=I$ or $II$, and $0<d_2<0.025$ for positivity of $d_0$.

For the sake of completeness, we document
the tensorial identities relevant to the derivation of the model
\beqy
&&\sum_{ab} {\bf c}_a^K {\bf c}_a^K = \sum_{ab} {\bf c}_b^K {\bf c}_b^K =
2 A_K \bI,  \nn\\
&&\sum_{ab} {\bf c}_a^K {\bf c}_b^K = 0,        \nn\\
&&\sum_{ab} (c_a^K)_i  (c_a^K)_j (c_a^K)_k (c_a^K)_l =
2 Z_K \GD_{ijkl} + 2 Y_K \Gd_{ijkl}, \\
&&\sum_{ab} (c_a^K)_i  (c_a^K)_j (c_a^K)_k (c_b^K)_l = 0, \nn\\
&&\sum_{ab} (c_a^K)_i  (c_b^K)_j (c_b^K)_k (c_b^K)_l = 0, \nn\\
&&\sum_{ab} (c_a^K)_i  (c_a^K)_j (c_b^K)_k (c_b^K)_l =
A_K^2 \Gd_{ij}\Gd_{kl} -2Z_K\GD_{ijkl}-2Y_K\Gd_{ijkl},\nn
\eeqy
where $\GD_{ijkl}= \Gd_{ij}\Gd_{kl}+ \Gd_{ik}\Gd_{jl} + \Gd_{il}\Gd_{jk}$,
$\Gd_{ijkl}=1$ only if $i=j=k=l$, otherwise is $0$;
$A_1=2$, $A_2=4$, $Z_1=0$, $Z_2=4$, $Y_1=2$, and
$Y_2=-8$. $I$ represents the identity matrix.

The viscosity and resistivity can be calculated using the Chapman-Enskog
expansion, in a similar fashion as it was done in Section III,
\beqy
\nu&=&\f{\tau+1}{6} \lab{nu17b}    \\
\mu&=&\f{3\tau-1}{6}.    \lab{mu17b}
\eeqy
The transport coefficients have been numerically
measured for the case of decaying shear flows, and the values found
were in agreement better than $0.1\%$ with the predictions of (\ref{nu17b})
and (\ref{mu17b}).


\section*{Figure Captions}
\begin{description}

\item Fig. 1
Profiles of $v_x$ vs. $y/L$ for Hartmann number
$H=0$, $1$, $3$, $5$, $8$, and  $13$, from
top to bottom. The $H=0$ (unmagnetized) case
corresponds to the Poiseuille flow. The lines
indicate analytical results, and the symbols
are the solutions provided by the LBE scheme.

\item Fig. 2
Profile of $B_x$ vs. $y/L$ for $H=3$ ($+$ symbol),
$H=1$ ($\diamond$ symbol), and $H=13$
($\bigtriangleup$ symbol). The lines indicate
analytical solutions, and the symbols represent
the LBE solutions. Simulations with
$H=5$ and $8$ were carried out with similar
results, and they are not presented here for clarity.

\item Fig. 3
Time histories of bulk quantities for the LBE run and the spectral run.
Evolution of a) the mean square vector potential
($A$); b) the total (magnetic plus kinetic)
energy $E$;
c)  kinetic energy ($E_k$); d)  mean square vorticity
(the enstrophy $\GO$), and e) the mean square current $J$.
The LBE run is indicated by the solid line and the spectral run by the
dashed line.

\item Fig. 4
Contours of constant magnetic field
(constant $a$), for the spectral (SP) and LBE runs, at times
approximately equal to one, seven, and ten.
Growth of magnetic islands can be observed for both runs.

\item Fig. 5
Lines of constant vorticity at times approximately
                        equal to one, seven, and ten, for both runs.
                        Quadrupole-like structures can be noticed
                         in the region of the $X$-points.

\item Fig. 6
Contours of constant stream function ($\psi$)
                        at times approximately equals to one, seven,
                        and ten. Similar features can be observed for
                        both spectral (SP) and LBE runs.

\item Fig. 7
Lines of constant current density at times
                       approximately one, seven, and ten, for the LBE
                       run and the spectral run. Diffusion from the
                       sheets area can be observed, as well as
                       filamentation in the $X$-point regions.

\item Fig. 8
Evolution of $B_L/B_\perp$ for the LBE run,
                        where $B_L$ is the longitudinal component of
                        the magnetic field. The higher initial value
                         is due to the unequal physical lengths in the
                         $x$- and $y$-directions, associated to the use
                         of the hexagonal lattice.
\end{description}

\end{document}